\documentstyle[11pt,newpasp,twoside]{article}
\index{instructions}
\index{guidelines}

\def\edcomment#1{\iffalse\marginpar{\raggedright\sl#1\/}\else\relax\fi}
\reversemarginpar

\begin{document}
\title{RX J1856.5-3754: A strange star with solid quark
surface?}
\author{X. L. Zhang$^1$,~~R. X. Xu$^2$,~~S. N. Zhang$^{1,3}$}

\affil{$^1$Physics Department, University of Alabama in
Huntsville, AL 35899\\
$^2$Schools of Physics, Peking University, Beijing 100871\\
$^3$Physics Department, Tsinghua University, Beijing 100084}

\begin{abstract}
Within the realms of the possibility of solid quark matter, we
fitted the 500ks Chandra LETG/HRC data for RX J1856.5-3754 with a
phenomenological spectral model, and found that electric
conductivity of quark matter on the stellar surface is about
$>1.2\times 10^{18}$ s$^{-1}$.
\end{abstract}

\noindent
{\bf 1. Introduction}

Neutron stars provide a unique opportunity for researchers who
are interested in obtaining experimental information of matter in
extremely high density, especially of the density-dominated quark
matter if a special kind of neutron stars, the so called strange
stars, exist.
However, how to identify strange stars is still a matter of great
challenges, and becomes a hot topic in astrophysics.

The featureless spectra of isolated ``neutron'' stars may be
evidence for {\em bare} strange stars (Xu 2002), but a definitive
conclusion on the nature of the compact objects cannot be reached
until theoretically calculated spectra of the bare quark surface
are known. However, due to the strong nonlinearity of quantum
chromodynamics, it is almost impossible to present a definitive
and accurate calculation of the density-dominated quark-gluon
plasma from the first principles. Nevertheless, it is suggested
that cold quark matter with extremely high baryon density could be
in a solid state (Xu 2003). We then try to fit the thermal X-ray
spectrum of the brightest isolated neutron star RX J1856.5-3754 in
this regime.

\vspace{2mm} \noindent
{\bf 2. Fitting the data and results}

In solid quark matter, the interactions between electrons
or between electrons and photons,
could be responsible to the thermal
photon radiation, which could be analogous to the radiation
of metals to some extent.
We will fit the thermal spectrum obtained from Chandra
LETG/HRC observations of RX J1856.5-3754 (total exposure time
about 500ks) with the metal emissivity (Born \& Wolf 1980) $
\psi(\nu, T)=\alpha(\nu)B(\nu, T)$, with $
\alpha(\nu)=1-(2\sigma/\nu+1-2\sqrt{\sigma/\nu})/(2\sigma/\nu+
1+2\sqrt{\sigma/\nu})$ and $B(\nu, T)$ the blackbody emissivity.

Chandra  LETG/HRC has performed five observations of RX
J1856.5-3754 in March 2000 and October 2001, with a total exposure
time $\sim$500~ks. Starting from the event2 files and following
the CIAO threads ({\em http://cxc.harvard.edu/
ciao/threads/gspec.html}), we extracted the spectra and generated
the corresponding effective area files for each observation, and
combined them for fitting.
The plus order and minus order spectra were combined seperately,
and a joint fitting was performed. To avoid the artificial features
across the gaps between CCDs, we ignored
0.18--0.21~keV in the plus order and 0.21--0.26~keV
in the minus order.
The spectra were grouped so that each channel has at lest
300~counts.
We fit the spectrum with both a pure black-body model and a metal
thermal model, both with interstellar absorption. The fitting
result is shown is Table~\ref{tab:fitting_result}.
\vspace{-5mm}
\begin{table}[ht]
\begin{center}
\small
\begin{tabular}{l|ccccc}
\hline\hline
 Model  & nH     & $kT_{\mathrm bb}^\infty$
 &  $ R_{\mathrm bb}^\infty $   & $\sigma$ & $\chi^2$/dof \\
 & \footnotesize $10^{20}{\mathrm cm}^{-1}$    & \footnotesize eV
 & \footnotesize km ($d$/120pc)      & \footnotesize $10^{18}{\mathrm s}^{-1}$ & \\
\hline
(a) & $0.86\pm0.03$ & $64.1\pm0.5$
    & $4.1\pm0.1$   & -    & 830/970 \\
(b) & $0.72\pm0.03$ & $59.5\pm0.5$ & {\footnotesize $>$(7.4, 5.9,
4.9)}
    & {\footnotesize$>$(1.24, 0.38, 0.14)} & 809/969 \\
\hline\hline
\end{tabular}
\caption{Fitting result: (a) absorbed blackbody, (b) absorbed
metal thermal spectrum. In (b), the lower limits of $\sigma$ and
$R_{\mathrm bb}^\infty $ are in $1\sigma$, $2\sigma$ and $3\sigma$
confidence levels, respectively.} \label{tab:fitting_result}
\end{center}
\end{table}

\vspace{-5mm} The metal model provides a slightly better fit to
the data. Since $\sigma$ and $R_{\mathrm bb}^\infty $ are coupled
together, we can only determine lower limits for the value of
$\sigma$ or $R_{\mathrm bb}^\infty $ from the fitting.
The parameters for the absorbed black-body model are a little
different from those obtained from the same observations by Burwitz et
al (2003). A possible reason is that we used different binning of
the spectra. However, the lower limit of neither $R_{\mathrm bb}^\infty $ nor
$\sigma$ is sensitive to the binning.

\vspace{2mm} \noindent
{\bf 3. Discussions}

Within the realms of solid quark surface, we have fitted the 500ks
Chandra LETG/HRC data for the brightest isolated neutron star RX
J1856.5-3754 with a phenomenological spectral model. However, the
UV-optical excess (e.g., Burwitz et al. 2003) has not been
included in the fitting.
Actually the origin of this excess is still academically
controversial. We suggest that the emission is not from the
stellar surface.
During a propeller phase with low accretion from interstellar
medium, a quasistatic atmosphere (envelop) may form around the
magnetopshere of RX J1856.5-3754. The dissipation of stellar
rotation energy may heat the envelop, which could be responsible
for the UV-optical emission.
Detailed investigation on this issue is necessary if one conceives
a study to know the environmental status of isolated neutron
stars.

If the electrons near the Fermi surface are responsible for the
conduction, the fitted $\sigma$ implies the relaxation time
$\tau>\sim 8\times 10^{-21}$ s, while the $e-e$ collision
timescale $\tau\sim 2.3\times 10^{-16}$ s.
The conductivity fitted could be reasonable.

Strong magnetic fields and/or fast rotation may help a neutron
star to reproduce a featureless spectrum. However, the thermal
radiation mechanism is very different from ours, which results in
a possible test in the future.

\acknowledgments This work is supported by NSFC (10273001) and the
Special Funds for Major State Basic Research Projects of China
(G2000077602).

\end{document}